\documentclass[conference]{IEEEtran}
\IEEEoverridecommandlockouts
\usepackage{cite}
\usepackage{amsmath,amssymb,amsfonts}
\usepackage{graphicx}
\usepackage{textcomp}
\def\BibTeX{{\rm B\kern-.05em{\sc i\kern-.025em b}\kern-.08em
    T\kern-.1667em\lower.7ex\hbox{E}\kern-.125emX}}
\usepackage[ruled]{algorithm2e} % For algorithms

\usepackage{amsmath}

\usepackage{graphicx}
\usepackage{caption}
\usepackage{hyperref}
\usepackage{epsfig, subfigure, amsmath, amssymb}
\usepackage{amsthm}
\usepackage{algpseudocode}
\usepackage{epstopdf}
\usepackage{color}
\usepackage{siunitx}
\usepackage{hhline}
\usepackage{comment}
\usepackage{makecell}
\usepackage{ragged2e}
\usepackage{array}

\definecolor{red}{rgb}{1,0,0}
\SetAlFnt{\small\color{black}\tt}
\LinesNumbered

\usepackage{multicol}

\begin{document}

\title{Improving DGA-Based Malicious Domain Classifiers for Malware Defense with Adversarial Machine Learning\\
%{\footnotesize \textsuperscript{*}Note: Sub-titles are not captured in Xplore and should not be used}

}

\author{\IEEEauthorblockN{Ibrahim Yilmaz, Ambareen Siraj, Denis Ulybyshev}
\IEEEauthorblockA{\textit{Department of Computer Science} \\
\textit{Tennessee Technological University}\\
Cookeville, USA\\
{yilmaz42, asiraj, dulybyshev@tntech.edu}}}
\maketitle

\begin{abstract}
Domain Generation Algorithms (DGAs) are used by adversaries to establish Command and Control (C\&C) server communications during cyber attacks. Blacklists of known/identified C\&C domains are often used as one of the defense mechanisms. However, since blacklists are static and generated by signature-based approaches, they can neither keep up nor detect never-seen-before malicious domain names. Due to this shortcoming of blacklist domain checking, machine learning algorithms have been used to address the problem to some extent. However, when training is performed with limited datasets, the algorithms are likely to fail in detecting new DGA variants. To mitigate this weakness, we successfully applied a DGA-based malicious domain classifier using the Long Short-Term Memory (LSTM) method with a novel feature engineering technique. Our model’s performance shows a higher level of accuracy compared to a previously reported model from prior research. Additionally, we propose a new method using adversarial machine learning to generate never-before-seen malware-related domain families that can be used to illustrate the shortcomings of machine learning algorithms in this regard. Next, we augment the training dataset with new samples such that it makes training of the machine learning models more effective in detecting never-before-seen malicious domain name variants. Finally, to protect blacklists of malicious domain names from disclosure and tampering, we devise secure data containers that store blacklists and guarantee their protection against adversarial access and modifications. 
\end{abstract}


\section{Introduction}
\begin{comment}

Malwares are often created by attackers to carry out malicious acts on or by compromised victim devices on a massive scale. To perform such malicious activities, a communication channel must be established between a victim machine and a C\&C server. Using these communication channels, botnets are remotely controlled by the attackers. Many different types of cyber-attacks can be performed through botnets, such as stealing users' sensitive information, sending spam, and denial-of-service attacks. According to a study, the number of exploited computers used in a botnet can be more than a million \cite {rossow2014amplification}, which demonstrates the level of danger and damage caused by these armies of zombie machines.
\end{comment}

\par Security researchers have developed many different defense mechanisms in order to protect computer systems against malware and malicious botnet C\&C communications. Blacklists of malicious websites are one of the most commonly used defense mechanisms where lists of domain names or IP addresses that are flagged as harmful are maintained. Any messages to/from these listed sites are feared to host potential C\&C servers and hence blocked to prevent any further communications. As a counterattack, attackers developed DGAs as a measure to thwart blacklist detection \cite{DGA}. 

\begin{figure*}
\centering
\includegraphics[width=16cm]{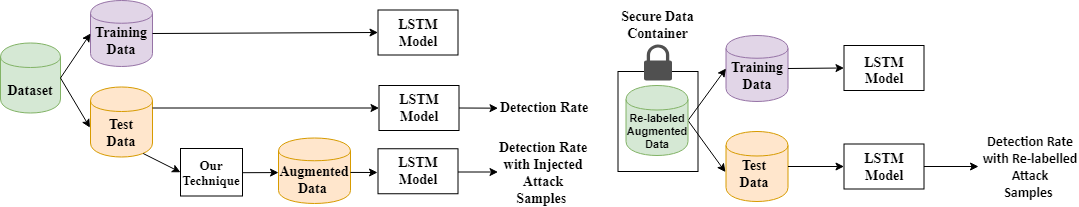}
\caption{Overview of the Proposed Methodology.}
\label{augmented}
\end{figure*}

\par In recent years, hacker communities have been utilizing DGAs as the primary mechanism to produce millions of malicious domain names automatically through pseudo-random domain names in a very short time period \cite {yadav2010detecting}. Subsets of these malicious domain names are utilized to map to the C\&C servers. These dynamically created domain names successfully evade static blacklist-checking mechanisms. Additionally, as one domain gets recognized and blocked, the C\&C server can easily switch to another one. 

\par To overcome the limitations of static domain blacklists, machine learning (ML) techniques have been developed to detect malicious domain names and these techniques have yielded mostly promising results \cite {woodbridge2016predicting}, \cite {yilmaz2020addressing}, \cite {yilmaz2020practical}. However, the ML models do not perform well with never-seen-before DGA families when an unrepresentative or imbalanced training dataset is used. To address this problem, we propose a novel approach to generate a rich set of training data representing malicious domain names using a data augmentation technique.

Data augmentation of an existing training dataset is one way to make ML models learn better and, as a result, perform more robustly. Nevertheless, classic data augmentation merely creates a restricted reasonable alternative. In our approach, as illustrated in Figure \ref{augmented}, an adversarial machine learning technique is used to generate a diverse set of augmented data by means of data perturbation. The generated adversarial domain names are extremely difficult to differentiate from benign domain names. As a result, the machine learning classifier misclassifies malicious domains as benign ones. Afterwards, these adversarial examples are correctly re-labeled as malicious and reintroduced into the existing training dataset (see Figure \ref{augmented}). In this way, we augment the blacklist with diverse data to effectively train the machine learning models and increase the robustness of the DGA classifiers.

In addition, we devise a secure container to store and transfer the blacklists of malicious domain names in encrypted form as a Protected Spreadsheet Container with Data (PROSPECD), presented in \cite{CBMS2020}. %, which ensures confidentiality of the blacklist and prevents any unauthorized modifications of it. 
PROSPECD provides confidentiality and integrity of the blacklists so that they can be used as training data to build a secure model. In addition to data integrity, PROSPECD provides origin integrity. This container protects the adversarial samples, used to teach the model, from unknown adversarial perturbations.  %uld An adversary might create new malicious domain names by using algorithms that are fundamentally similar to ours to fool the autonomous blacklist DGA detector. However, it would be computationally infeasible for an adversary to spoof the fake blacklist origin because of the RSA-based digital signature used to sign PROSPECD. Furthermore, our approach can detect several types of blacklist data leakages made by authorized entities to unauthorized ones.
The protected blacklist can be marketed commercially \cite{profit} to internet service providers and companies who need to maintain their own internal blacklists.

\subsection{Our Contributions}

Blacklist is a security strategy that keeps network flow and computer environments secure \cite{Blacklists}. Typical network traffic blacklists include malicious IP addresses or domain names, which are blocked from communication attempts in both directions. However, the coverage of blacklists is insufficient and unreliable because adversarial hacker communities can compromise a system by generating malware domains dynamically using DGAs that easily bypass the static blacklist. Kuhrer et al. \cite {kuhrer2014paint} evaluated 15 public malware blacklists as well as 4 blacklists, served by antivirus vendors. Their findings show that blacklists fail to protect systems against prevalent malware sites generated by DGAs.

In order to address the shortcomings of the above problem, researchers have mostly proposed solutions based on reverse engineering techniques to identify and block bot malware \cite{ligh2010malware}. However, such solutions are not always feasible due to the obfuscation of underlying algorithms, as hackers adapt their algorithms swiftly to exploit the vulnerabilities in the system.
Other alternative solutions require auxiliary contextual information. One of these alternative solutions focuses on the network traffic analysis \cite {zhang2011detecting}, \cite {yen2010your} or broad network packet examination \cite {manni2015network}, \cite{aziz2018electronic}. However, these techniques may not be able to keep up with large-scale network traffic. Therefore, there is a need for a sophisticated network traffic analysis tool for effective blacklisting.

In response to the above issues, detection of malicious
domains has increasingly evolved towards usage of machine learning techniques. Performance of the solutions proposed to automatically detect malicious domains mostly suffers from never-seen-before malicious domains. This is due to the lack of generalization when a model is not trained effectively with representative or balanced training dataset. For this reason, blacklists must be constantly updated in order to identify and prevent DGA generated malicious domain connections.

Data augmentation is an approach where more data is created from existing data in a way that can enhance the usefulness of the application. Anderson et al. \cite{anderson2016deepdga} demonstrated how data can be augmented more effectively by using an adversarial machine learning technique. The researchers enhanced adversarial domain names using Generative Adversarial Network (GAN) methodology. In their approach, two neural networks are trained simultaneously, and later trained with a dataset which includes adversarial samples to harden the DGA classifier. However, the main drawback of this approach is unpredictability of desirable results because of difficulties of controlling both classifiers at the same time, even though a good optimization algorithm is used. As a result, it fails to always converge to a point of equilibrium to generate new domain names. Additionally, controlling the diversity of produced samples is challenging with GAN models \cite {berthelot2017began},  \cite{yilmaz2019expansion}. In such cases, the newly generated data do not add to the diversity of the current data. Hence, this solution alone cannot increase the malicious detection capabilities of blacklists against never-before-seen DGA families.

To improve the accuracy of such detection mechanisms, we propose a new technique based on data perturbation without relying on a fresh public blacklist or external reputation database. In our approach, we observe how the model works and use the knowledge to mislead the DGA classifier. To do this, a noise, carefully calculated from the observation is added to the DGA based malicious domains to appear non-malicious. These adversarial samples are then predicted as benign by the machine learning (ML) model. Such adversarial attack can be addressed with an adversarial training  \cite{goodfellow2014explaining}. Therefore, after correctly labeling these seemingly benign adversarial samples, the ML model is trained with the augmented dataset. The experimental results demonstrate that the retrained ML model is able to detect never-before-seen DGA malwares better than any other similar approaches. 

Our work has the following contributions:
\begin{itemize}
\item Using a machine learning technique based on the Long Short-Term Memory (LSTM) model for automatic detection of malicious domains using a DGA classifier that analyses a massive labeled DGA dataset, character by character over.  
\item \textit{To the best of our knowledge} it is the first study to propose the generation of malicious domain names using a data perturbation approach in order to expand the training dataset with adversarial samples.
\item Demonstrating that, as expected, the LSTM model fails to recognize newly introduced adversarial samples in augmented training dataset.
\item Applying an adversarial training to train the model with correct labelling of the adversarial samples in the training dataset to increase the model's generalization ability. 
\item Demonstrating that the augmented training dataset can help the LSTM model to detect not only never-seen-before DGAs, but also novel DGA families.
%\item Secure data container stores detected malicious domain names in an encrypted form as a spreadsheet file. Only authorized entities can access the blacklisted domain names. Each separate data worksheet is encrypted with a separate symmetric encryption key, generated on-the-fly, using the novel key generation scheme. As a result, the key is not stored inside the container or on any Trusted Third Party (TTP), which makes the scheme more secure.    
\end{itemize}

\par The rest of this paper is organized as follows: the literature review in the context of our work is discussed in section \ref{related}. The necessary background for DGA-based malicious domain models is reviewed in Section \ref{background}. The core design of our system, including the adversarial machine learning models to generate malicious domain names and data containers to store them, are presented in Section \ref{design}. We discuss the implementation of the adversarial machine learning models in Section \ref{implementation}. The evaluation results of our study are presented in Section \ref{evaluation}. Section \ref{conclusion} concludes the paper.

\section{Related Work}
\label{related}
\par Domain Generation Algorithms and detection of malicious domain names have been previously analyzed by different researchers for a number of years. Daniel et al. \cite{plohmann2016comprehensive} presented a taxonomy of DGA types by analyzing characteristics of 43 different DGA-based malware families and compared the properties of these families. They also implemented previous studies with 18 million DGA domain data that was created to identify malicious domains. They reported further progress in DGA detection. 

Detection of DGA botnets became feasible with the implementation of powerful machine learning models. Lison et al. \cite{lison2017automatic} implemented a recurrent neural network model for the detection of DGAs. Their empirical study detected malicious domain names with high accuracy. Justin et al. \cite{ma2009beyond} defined several models to detect malicious web sites, including a logistic regression, support vector machine, and bayesian model. They used the DMOZ dataset for benign websites while PhishTank and Spamscatter were used for malicious websites. Bin et al. \cite{yu2018character} addressed the same issue using different machine learning classifiers for the detection of DGAs. They created a convolutional neural network (CNN) and a recurrent neural network (RNN) machine learning model for the classification of malicious and benign domain names. They compared the results of both models in terms of their performance and reported that both models performed comparably. Duc et al. \cite{tran2018lstm} dealt with the multiclass imbalance problem of LSTM algorithms for the detection of malicious domain names by generating DGAs. The authors claimed that the LSTM algorithms performed poorly with imbalanced datasets. To tackle this imbalanced dataset problem, they proposed a Long Short-Term Memory Multiclass Imbalance (LSTM.MI) algorithm and showed that their proposed algorithm provided more accurate results by implementing different case studies.

\par In addition, Mayana et al. \cite{pereira2018dictionary} introduced a WordGraph method to recognize dictionary based malicious domains. The authors asserted that more sophisticated DGAs are able to avoid detection of conventional machine learning classifiers. They carried out their experiments by extracting dictionary information without using reverse engineering. Bin et al. \cite{yu2017inline} defined a deep neural network model as an inline DGA detector. They caution about most of the available datasets not being good representations for malicious domains or outdated. Hence, machine learning models' perform poorly when trained using such datasets. Furthermore, they explained that reverse engineering was a difficult method for training models. To tackle these problems, the researchers offered a novel detector for malicious domains without the need for reverse engineering. Their proposed technique was based on real traffic and reported to detect malicious domains in real time. Woodbridge et al. \cite{woodbridge2016predicting} created a machine learning classifier based on the LSTM network to detect malicious domain names in real time. Their classifier detected multiclass domains by categorizing them into particular malware families. The model predicted a domain name as malicious or benign according to this domain name without any additional information.

However, although these studies achieved high detection rates for particular DGA families, the performance of machine learning based detection systems are poor with new DGA variants when the models are trained with unrepresentative or imbalanced training datasets. To handle this issue, Anderson et al. \cite{anderson2016deepdga} offered a GAN algorithm to generate new domain names. In their GAN approach, they implemented two different deep neural network classifiers named discriminator and generator. According to this GAN methodology, new malicious domain names are generated by the generator, which evades the discriminator's detection. Their case studies demonstrated that new malicious domain names also bypass a random forest classifier. Once the model was trained with adversarial samples, it hardened the model against new DGA families. However, the authors did not test it on DGA families created using a dictionary. Additionally, implementation of this approach is challenging due to the need of controlling two machine learning models, which might be unsuitable for detecting malware-related domain names. Unlike this approach, we propose to augment data by using an efficient data perturbation technique that generates hard-to-detect DGA families and identifies DGA types that are created either randomly or using a dictionary. 

To overcome limitations of machine learning models in aforementioned circumstances, Curtin et al. \cite{curtin2019detecting} proposed to combine a neural network model with domain registration supplementary information. This additional information, which is known as WHOIS data, helped the neural network model to efficiently identify the most difficult samples that were generated using English words. However, cybercriminals take advantage of bulk registration services by registering thousands of domain names in a short time, several months before the start of nefarious activities \cite{zhauniarovich2018survey}. In addition, unauthorized people can access this information and falsify it by impersonating as legitimate users. This makes the information questionable. Compared to this, our approach efficiently detects DGA families solely based on the domain names, without relying on any supplementary information.

PROSPECD container used to store and transfer blacklisted malicious domain names, is presented in \cite{CBMS2020}. Compared to the privacy-preserving data dissemination concept as proposed by Lilien and Bhargava \cite{lilien2006}, it has the following features:
\begin{itemize}
    \item Detection of several types of data leakages that can be made by authorized entities to unauthorized ones;
    \item  Enforcement of access control policies either on a central server or locally on a client's side in a Microsoft Excel\textsuperscript{\textregistered}\footnote{This paper is an independent publication and is neither affiliated with, nor authorized, sponsored, or approved by, Microsoft Corporation \cite{microsofttrademark} } Add-in or in a cross-platform application \cite{CBMS2020}.
    \item Container implementation as a digitally signed %\cite{ExcelDigitalSign} 
    watermarked Microsoft Excel\textsuperscript{\textregistered}-compatible spreadsheet file with hidden and encrypted data and access control policies worksheets. %uld , including an encrypted metadata worksheet with access control policies.
    \item On-the-fly key derivation mechanism for data worksheets. % June 16, which relies on the hash values of inputs, including the hash of  Authentication Server (AS)'s private key. This information can be retrieved from an AS only by authorized entities whose identity, represented in X.509 certificate, has been verified by AS.   
    %\item A different policy enforcement mechanism: policies are enforced either locally in a Microsoft Excel\textsuperscript{\textregistered} Add-in \cite{CBMS2020}, or on a back-end server that hosts the spreadsheet file.
\end{itemize} 
The primary difference between PROSPECD and an Active Bundle \cite{othmane2010thesis}, \cite{othmane2009}, \cite{ranchal2015cross} is that PROSPECD does not store an embedded policy enforcement engine (Virtual Machine). 

In contrast with a solution to encrypt the desired cells in a spreadsheet file, proposed by Tun and Mya in \cite{tun2010}, in PROSPECD all the data worksheets are encrypted with the separate keys, generated on-the-fly. PROSPECD supports role-based and attribute-based access control. %uld , using the access control policies, stored in encrypted form as a separate worksheet.
Furthermore, digital and visual watermarks are embedded in PROSPECD to enable detection of data leakages. A Secure Data Container, proposed in \cite{secureICS} to store device state information, only supports centralized policy enforcement. %uld on a Trusted Third Party. 
PROSPECD supports both centralized and local policy enforcement mechanisms \cite{CBMS2020}. %, which is important for environments with limited Internet connectivity  Furthermore, our approach provides origin integrity, in addition to data integrity, which is essential for storing blacklists of malicious domain names.

\section{Background}
\label{background}
In this section, we review background information related to  our research.

\subsubsection{\textbf{Domain Generation Algorithm (DGA)}} 
Domain generation algorithms are primary means to connect various families of malware with new or never-before-seen domains to avoid detection. There are many such DGA-based malware families (malware that connect to DGA generated domain names). According to a study, the five most known families are \textit{Conficker}, \textit{Murofet}, \textit{BankPatch}, \textit{Bonnana}, and \textit{Bobax} \cite{top}. Although many DGA-based domain names are produced randomly, some are generated using a dictionary. The detection of these types of domain names is more difficult because of their similarity to legitimate domains.

\subsubsection{\textbf{Gradient Descent Algorithm}} 

The gradient descent algorithm is the most popular optimization method for using machine learning classifiers to minimize errors. %It is an algorithm based on the first derivative of a cost function% 
It takes into account the first derivative when modifying all parameters under considerations \cite{gradient}. Gradient descent always strives to find the most appropriate way to minimize errors. The learning process starts with randomly producing weight values. Most of the time, these values are set to an initial value of zero and are used to calculate the lost function value. It then uses the gradient descent algorithm to find a way to reduce the lost function. All weights are updated through the backpropagation process based on the gradient descent algorithm. We generate new adversarial domain names in our data augmentation method by utilizing and modifying the gradient descent algorithm's behavior. In Section \ref{implementation}, explanation of how such adversarial samples are created is discussed in detail. %as summarized below.

%\begin{equation}\nonumber
%\begin{split}
%\mathit{W}_{final} = \mathit{W}_{current} - \alpha * \dfrac{\mathit{dJ(W)}}{\mathit{dw}}
%\end{split}
%\end{equation}

%Here $\alpha$ represents the learning rate and $\dfrac{\mathit{dJ(W)}}{\mathit{dw}}$ represents the derivative of the lost function in terms of weights. $\mathit{W}_{current}$ is the current position of the particular weight and $\mathit{W}_{final}$ is the updated version of this weight.

\subsubsection{\textbf{Long Short-Term Memory (LSTM) Model}} 

The LSTM model, a specialized Recurrent Neural Network (RNN), is used in our approach for automatic detection of malicious domains. RNNs are known as supervised machine learning models that are commonly used to handle the processing of sequential data \cite {hochreiter1997long}. An RNN takes the previous and current inputs into account while the traditional networks consider all inputs independently. In our study, we implement the model on a character by character basis, so that it captures the importance of the order of the characters' occurrence in the word. Essentially, the model learns the occurrences of the characters in a sequential way. For example, for the domain name \textit{google}, without the top-level domain, the model first learns the character \textit{g}, then \textit{o}, predicting that it succeeds \textit{g}. However, traditional neural networks do not take the position of the characters into account. In addition, the traditional neural networks process fixed-sized input. Similarly, they can create fixed-size output, whereas RNN does not have such limitations.

%%A conventional RNN takes input \textit{X} as an embedding vector of the first character. It feeds into the model and computes the predicted output \textit{Y} along with updating the internal state of \textit{h}. The information is sent from the current cell to the next cell systematically. This process is shown below with the following equation \cite{rnn}:
%\begin{equation}\nonumber
%h_{t} =f(h_{t-1}, X_{t)}
%\end{equation}
%Where $h_{t}$ and $h_{t-1}$ are the current state and the previous state respectively and $X_{t}$ denotes the input at the time t.

When input data like domain names have long term sequences, traditional RNN struggles to learn such long data dependencies, which is known as the vanishing gradient problem  \cite{hochreiter1998vanishing}. In order to avoid this problem, LSTM, which is a special kind of RNN, was introduced in \cite{hochreiter1997long}.

LSTM relies on the gating mechanisms, where the information can be read, written or erased via a set of programmable gates. This allows recurrent nets to keep track of information over many time-steps and gain the ability to preserve long term dependencies. For example, \textit{47faeb4f1b75a48499ba14e9b1cd895a} is a malicious domain name, which has a 32 character length. LSTM keeps track of the relevant information about these characters throughout the process. Furthermore, a study has shown LSTMs outperform previous RNNs for the solution of both context-free language (CFL) and regular language problems \cite{gers2001lstm}. Researchers have reported that LSTMs generalize better and faster, leading to the creation of more effective models. In our study, the LSTM model learns how to detect DGA-based malicious domains in a similar way to what is mentioned above.

%LSTM has three different components, which allows it to learn effectively. These are: \textit{short term memory}, the\textit{ previous cell}, and the \textit{long term memory}. The long term memory is known as the cell state, whereas the short term memory is known as the hidden state. LSTM consists of three gates called \textit{input gate}, \textit{forget gate}, and \textit{output gate} \cite{sak2014long}. The input gate chooses what new information will be kept in the long term memory, whereas the forget gate decides whether to keep it or not. The output gate takes into account the current input, the previous short term memory, and the calculated long term memory in order to generate the new short term memory that will be sent to the cell in the next step. 

\section{Core Design}
\label{design}

\subsection{Generating New Malicious Domain Names}

Machine learning models' performances substantially rely on the training dataset crucial for building effective classifiers. However, one of the biggest challenges with any ML model is accumulating a training dataset that is representative and balanced enough to enable the creation of an effective machine learning model. This process might be costly or time-consuming or both. A restrictive training dataset can lead to poor performance of the ML model and that is the primary reason DGA classifiers do not work well with automated malicious domain name detection. With traditional blacklists used in training, ML classifiers cannot detect never-before-seen DGA families. The model needs to be readjusted constantly with new variations of training data for effective threat  detection. To address this issue, we propose to create a blacklist with domain names using a novel adversarial machine learning technique. 

Our adversarial approach is based on data perturbation techniques inspired by \cite{goodfellow2014explaining}, where domain names are influenced based on the gradient descent algorithm of a targeted DGA classifier with regards to the classifier loss. Changing the gradient of the classifier maximizes loss function instead of minimizing it can mislead the malware detection classifiers. Even though these domains are malicious, the DGA classifier, based on the LSTM model, predicts them as benign. As a result, new adversarial domain are generated that can seem benign by not matching the blacklist data. Our method is clarified mathematically below \cite{goodfellow2014explaining}.

 Let \textit{x} be a given malicious domain name, \textit{y} be labeled as malicious, \textit{M} be a DGA classifier that M(x): x $\rightarrow$ y, \textit{$\hat{x}$} is crafted domain name using our adversarial attack and \textit{$\hat{y}$} is the class label that M(x): $\hat{x}$ $\rightarrow$ $\hat{y}$
\begin{equation}
 \hspace{0.3cm} objective \hspace{0.6cm}  max \hspace{0.1cm}l(M,\hat{x},y)
\end{equation}
\begin{equation}
\hspace{1cm} y \neq  \hat{y}
\end{equation}
\begin{equation}
subject \hspace{0.1cm} to \hspace{0.5cm} \hat{x} = x + \delta x 
\end{equation}

Here \textit{l(M.x,y)} is the loss function of the DGA classifier in (1). A newly created adversarial domain name is predicted as benign by the DGA classifier in (2). \textit{$\delta$x} in (3) represents perturbation added to a vector form of the given domain name. $\delta$x is calculated below \cite{goodfellow2014explaining}:
\begin{equation}
\hspace{0.3cm} \delta x = \epsilon \hspace{0.1cm} sign(\nabla x \hspace{0.1cm}l(M,x,y))
\end{equation}

Here \textit{sign} ($\nabla$x l(M,x,y) ) represents the direction of the loss function, which minimizes the loss function of the DGA classifier, and $\epsilon$ controls the expansion of the noise in (4). The smaller epsilon value perturbs the original feature vector slightly, while the larger one perturbs the original feature vector significantly. This misleads the DGA classifier to a great extent. On the other hand, a larger perturbation can be more easily detected than a smaller one by human eyes.  

The calculated noise is included in each character embedding of input data. The resulting embedding characters with noisy calculations are compared to each character's embedding using \textit{cosine similarity} \cite {cosine_similarity} to measure each distance. The final character is chosen based on this operation. The design of the model is demonstrated in Figure 2. According to the example in the figure, character g turns to c with our use of adversarial learning. Our adversarial domain name generation algorithm is summarized in Algorithm 1.

\begin{figure}[!ht]
\centering
\includegraphics[width=8cm]{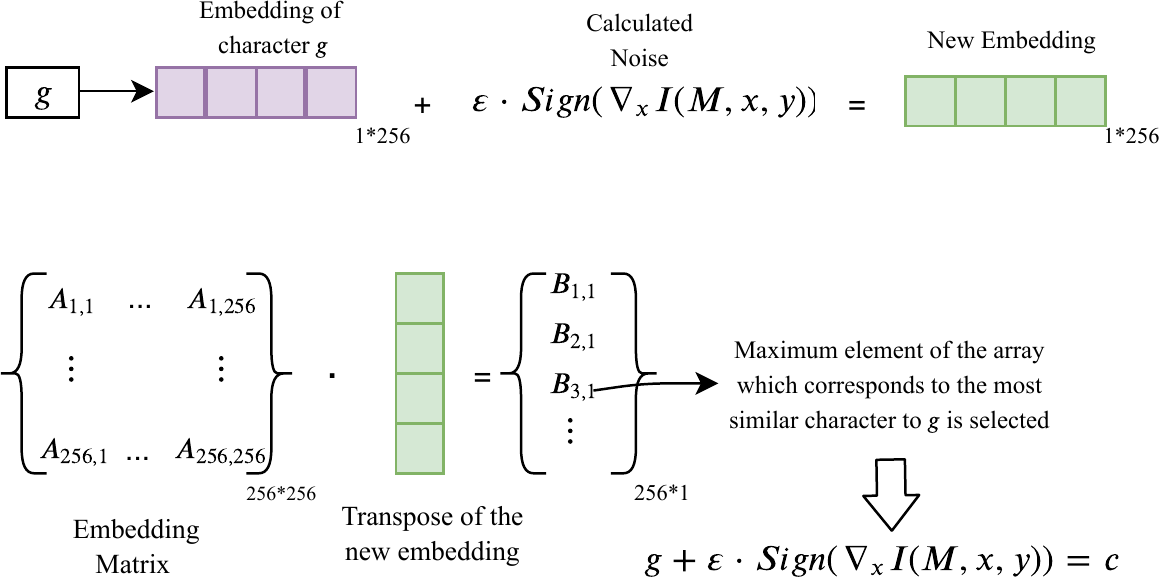}
\caption{Generation of Adversarial Samples Using Character-Level Transformations.}
\label{DDOS}
\end{figure}

\begin{algorithm}
\caption{Pseudocode of our proposed \textit{adversarial domain name generation} approach}
\label{alg:generator}
\SetKwProg{generate}{Function \emph{generate}}{}{end}
\noindent \textbf{Input:} 
\\- Train data pair $\{X_{i},Y_{i}\}$ where $X_{i}$ = Each domain name and $Y_{i}$ = Corresponding ground-truth label\\
-    X =($x_{1}$ || $x_{2}$ ||...|| $x_{n}$) where $x_{i}$ = Each character of a given input domain name\\
- Training iteration number $N_{itr}$, Number of adversarial examples $N_{adv}$, Number of training samples $N_{train}$, Number of character of a given input domain name $N_{total}$\\
- Test data pair $\{X_{j},Y_{j}\}$\\
\generate{domain names}{
     \For {iteration = 0, ..., $N_{itr}$}{
     \indent \indent Update all parameters based on gradient descent \newline 
     } 
     \For {iteration = 0, ..., $N_{adv}$}{
          \For {iteration = 0, ..., $N_{total}$}{ \vspace{2mm}
     \indent \indent $\delta_{x_{i}}$ = $\epsilon \times sign (\nabla_{x_{i}}l(M,x_{i},y_{i}))$  \newline
     \indent  \# Calculate penetration for each character \vspace{2mm} \newline
\indent \indent ${z_{i}}$= $x_{i}$ + $\delta_{x_{i}}$

\vspace{2mm}
\indent \indent$ 
\underset{max} {\hat{x}}= \frac{ {z_{i}}  \bullet    x_{}} {\sqrt{{z_{i}} ^ 2 \times  {x_{i}} ^ 2}}
\vspace{2mm}
$ \newline 
\indent \# New characters are generated through our proposed method\newline
     
     }
     
    \noindent \textbf{Output:} $\hat{X}$ =($\hat{x}_{1}$ || $\hat{x}_{2}$ ||...|| $\hat{x}_{n}$) \newline
    \indent \# New malicious samples are generated\newline
    
}
}
\end{algorithm}

The DGA detectors can be seen as black-box devices in real-world settings. Since, in the black-box scenario, an adversary does not have any knowledge about the inner workings of the target model. Nevertheless, for the sake of simplicity, we implement our proposed technique under the white-box assumption, where we obtain optimum perturbation by accessing the target model so that we can compute gradients. Although the black-box assumptions can be perceived as more realistic for this work, it is important to keep in mind that previous studies showed that adversarial samples have the \textit{transferarability property} \cite {papernot2016transferability}. This means that an adversarial example generated for one DGA model is more likely to be misclassified by another DGA detector as well, since when different ML models are trained with the similar  dataset from the same source, they learn similar decision boundaries. We leave testing adversarial examples the under black-box settings for future work. %We hope that our method establishes a strong baseline for further research. %In this research study, we have shown successful generation of  adversarial domains using data perturbation technique under the white-box assumption.% W
%*\textbf{Discussion.} Goodfellow et al. \cite{goodfellow2014explaining}, who generated adversarial image samples by finding optimum noise, mentioned that the subtle noise additions do not change the actual class. It solely alters the model decision mechanism by deceiving the model. However, if the noise we add into the benign domain names changes the domain name even by one character, then we can confirm that the actual benign domain class has converted into a malicious one. For example, in Table \ref{transformation}, the domain name adobe turns into adobu by adding 0.6 penetration coefficient. In our study, we consider the effect of this situation by checking whether there is a change in the benign domain or not.

\subsection{Protected Spreadsheet Container with Data (PROSPECD) for Domain Names Blacklists}

We propose to use a PROSPECD data container, presented in \cite{CBMS2020}, to securely store and transfer blacklisted malicious domain names. In our use case, PROSPECD, implemented as an encrypted and digitally signed spreadsheet file, contains the following watermarked data worksheets: 

\begin{itemize}
\item{"Domain Blacklist" to store encrypted malicious domain names, detected by our classifier;}
\item{"Metadata" to store encrypted metadata, which include access control policies;}
\item{"General Info" to store encrypted information about the classifier used to detect the malicious domain names and its execution details.} %uld , time and network node information of the classifier's execution.}
\end{itemize}
 %from textcolor{red}

PROSPECD provides data confidentiality and integrity, origin integrity, role-based and attribute-based access control and centralized and decentralized enforcement of access control policies. Digital and visual watermarks, embedded into a PROSPECD spreadsheet file, enable detection of several types of data leakages that can be made behind-the-scenes by authorized parties to unauthorized ones \cite{CBMS2020}.   \\

\textit{\textbf{PROSPECD Generator.}}
The malicious domain names classifier runs on a trusted server. Once the blacklist of domain names is generated, the dedicated process writes it, as well as the relevant information, %on what classifier detected the malicious domain names, when and where, 
in a spreadsheet file. Then the PROSPECD generator, currently implemented as a command line utility, is called. It takes a spreadsheet file with the "Domain Blacklist" worksheet and two other worksheets ("General Info" and "Metadata") in a plaintext form, as an input, and generates a separate spreadsheet file with encrypted worksheets. Each separate worksheet is encrypted with a separate symmetric 256-bit AES key, generated on-the-fly \cite{CBMS2020}. 
%AES key generator takes three hash values as inputs: hash value of AS's private key, worksheet name and metadata worksheet. 
%15 June To get the hash value of AS's private  
%15 June It allows protection of data worksheets from unauthorized accesses since the AS only reveals hash value of its private key to known authorized entities. We rely on X.509 certificates for identity management.

\textit{\textbf{PROSPECD Data Access on a Trusted Server.}}
The PROSPECD container, stored on a trusted %back-end  
server, can be accessed remotely from a web viewer. The client opens the Authentication Server (AS)'s URL in a web browser, selects the data subset to retrieve (”Domain Blacklist”, "General Info" or "All") and enters their credentials: username (role) and password. The accessible worksheets from PROSPECD are decrypted, using the on-the-fly AES key derivation scheme, based on the client's role and attributes. These attributes include versions of a web browser and an operating system, as well as the type of device the client uses. Decrypted worksheets are sent to the client as a JSON object via https communication channel \cite{CBMS2020}. %uld Details can be found in \cite{CBMS2020}. 
PROSPECD supports %data requests via 
RESTful API. Table \ref{access-control-table} shows the access control policies. The role "User" can only access blacklisted domain names from the "Domain Blacklist" worksheet. % June 16 The "User" cannot access the "General Info" worksheet which contains information when the malicious domain name classifier worked and what detection algorithm was used. Also, the "User" role cannot access the "Metadata" worksheet that contains the access control policies.
The role "Administrator" is allowed to access all the worksheets and also to download the PROSPECD file from the server to their local device, to access the data locally or transfer it to other parties. %uld PROSPECD guarantees that a subject cannot access %extract and decrypt 

\begin{table}[!h]
\centering
\caption{PROSPECD Access Control Policies}
\sisetup{per-mode=symbol}
%\renewcommand\arraystretch{0.95}
%\setlength{\doublerulesep}{5pt}
%\rowcolors{3}{red!30}{cyan!5}

\begin{tabular}{||c c c c||} 
 \hline
     & Domain Blacklist & General Info & Metadata \\ [0.5ex] 
 \hline\hline
 Administrator & YES & YES & YES \\ 
 \hline
 User & YES & NO & NO \\
 \hline
 \end{tabular}
\label{access-control-table}
\end{table}

\textit{\textbf{Local PROSPECD Data Access.}}
Authorized parties can access PROSPECD data locally, either from the Microsoft Excel\textsuperscript{\textregistered} Add-in or from the standalone cross-platform application.
For the first option, a user needs to "download and install the Microsoft\textsuperscript{\textregistered} Excel Add-in" \cite{CBMS2020}, written in C\#\textsuperscript{\textregistered}, open a PROSPECD container and enter valid credentials. Then the PROSPECD's digital signature is verified. If it is valid, the "Metadata" worksheet is decrypted and access control policies, stored in this worksheet and shown in Table \ref{access-control-table}, are evaluated. Then the decryption keys for accessible data worksheets are derived \cite{CBMS2020}.  %If this is the first time the Add-in accesses the PROSPECD file, 
%uld "HTTPS request is sent to an AS to retrieve the hash value of its private key, which is needed to generate the right decryption key." \cite{CBMS2020}  %Then, the hash is stored in an encrypted configuration file for future use." 
 % This enables decentralized access to PROSPECD data, which is important for limited Internet connectivity environments. %It eliminates the necessity for the Add-in to contact an AS each time the Add-in is launched. 
To prevent unauthorized data disclosures, the authenticated user is not allowed to print or save the opened spreadsheet file once the Add-in has been launched. %Once the user is finished with their session they may close the application. 
When a user closes an application, all the data are encrypted back to their original values and the visibility of all the worksheets is reset back to VeryHidden\textsuperscript{\textregistered}, after which the application closes.  %For future uses, this hash value is encrypted and stored locally in the configuration file.
 %Then, the configuration file is encrypted.
%Each user within a given role is only allowed to see the data worksheets that they are allowed to access, based on access control policies evaluation made by the Add-in. Access control policies are identical to those on the trusted back-end server and are illustrated on Fig.~\ref{access-control-matrix}. The Add-in also disallows file saving during operation. This helps to prevent data leakage. 
%Each role has the same permissions for both supported policy enforcement mechanisms: in the trusted back-end server and the decentralized Add-in, which also disallows file saving during operation. This helps to prevent data leakage.

In addition to the Microsoft\textsuperscript{\textregistered} Excel Add-in, a cross-platform application was developed to view PROSPECD data \cite{CBMS2020}. %in a Read-Only mode \cite{CBMS2020}. 
This application provides a graphical user interface and does not allow the user to store the decrypted PROSPECD files locally, to prevent possible data leakages.

\section{Experimental Methodology}
\label{implementation}

This section describes the dataset used to build a DGA classifier based on an LSTM model, along with the explanation of the model implementation. 
%In addition, we cover how to generate malicious domain names in an evasive manner to augment a training dataset that can then be used for more robust machine learning models. The evaluation of the DGA model and the performance of the data perturbation approach are discussed in details in the Evaluation section.

\subsection{Dataset}
The experimental dataset includes one non-DGA (benign domain names) and 68 DGA families (malicious domain names). Data is collected from two different sources, which are publicly available \cite{majestic}, \cite{plohmann2016comprehensive}.

For benign domains, we use the Majestic top 1 million dataset \cite{majestic}. This dataset includes the top one million website domain names all over the world and the dataset is updated daily. For malicious domains, we obtain data from the DGArchive, which is a web repository of DGA-based malware families  \cite{plohmann2016comprehensive}. This repository has over 18 million DGA domains. We have worked on 68 DGA malware families with some being generated by traditional DGAs. The remaining families were produced by dictionary DGAs. We used both traditional-based DGAs and dictionary-based DGAs with over a hundred thousand malicious domains. To ensure a fair comparison, we used a subset of 135K samples from the Majestic top 1 million dataset so that the classifier does not bias towards the majority class and thus, prevent occurrence of overfitting.

% It is arranged by the number of referring subnets. It has 12 different features named: \textit{GlobalRank}, \textit{TLDRank},\textit{ Domain}, \textit{TLD}, \textit{RefSubNets}, \textit{RefIPs}, \textit{IDN\_Domain}, \textit{IDN\_TLD}, \textit{PrevGlobalRank}, \textit{PrevTldRank}, \textit{PrevRefSubNets}, and \textit{PrevRefIPs}. We implement a character-based detection to build a DGA classifier and therefore only consider domain names without relying on other features. 

\subsection{ML Model Implementation}

We implement our LSTM model in Python using Pytorch \cite{pytorch}. We use the LSTM at character level with application of character embeddings (vector forms of the characters). This means that we every character is encoded to a representative vector. We convert from the word spaces to vector spaces to extract better features for the machine learning classifier. Each ASCII character (total number is 256) represents a vector whose size is set to 256. In this way, we create a 256 by 256 embedding matrix where each row represents a character and character embeddings are represented by column vectors in the embedding matrix. Once the perturbation technique is employed, embedded vectors of the character of malicious domains are transformed into word space again.

We divided the dataset into training and test data. We use 90\% of the dataset for training and remaining is reserved for testing. The training is used by the model to learn detection of DGA-based malicious domains. In our implementation, only the domain names are considered by the model, and the characters are pulled from the domain names character by character. At each time interval, one character's corresponding vector is fed into the LSTM model. The character embeddings are randomly initialized at the beginning. The model is able to learn through the dependencies that is has with each other and the conditional probabilities of the aforementioned characters. Thus, each character's embedding is learned by the LSTM model itself and the matrix is filled with these embeddings. In the test phase, the unseen data is predicted by the model as malicious or benign. The model's performance is analyzed in section \ref{evaluation} in detail. 

In this work, our main goal is to augment the training dataset to increase the model's resiliency and improve performance for detection of never-before-seen or yet-to-be-observed DGA families. To do this, an optimally calculated noise is added to each character embedding of the input data by the data perturbation technique. The newly created embedding with the addition of noise may not be assigned to any character. Therefore, the model looks for the closest embedding character and assigns it as the character to that corresponding embedding. Here we use \textit{approximate similarity search} \cite {patella2009approximate} by applying c\textit{osine similarity} \cite {cosine_similarity} that takes dot products between newly created embedding and each row of the embedding matrix to calculate the similarity. The new character is assigned to the row matrix's corresponding character, yielding maximum similarity value.  

Technically speaking, the LSTM model consists of two hidden layers along with the input and output layers. The \textit{drop out}, a known regularization technique, is used with a rate of 0.5 in order to avoid overfitting. The fine-tuned parameters are found by using a batch size of 128 and a learning rate of 0.001 along with an epoch number of 6. To achieve learning rates lower than this, more iterations may be needed.

In addition, we use the \textit{adam optimization algorithm} 
%\cite {kingma2014adam}, 
an extension to stochastic gradient descent, to minimize the error by adjusting network hyper parameters in an iterative way. Furthermore, \textit{binary cross entropy} 
%\cite {de2005tutorial}, 
utilized for binary classification, is used as the loss function in order to measure the cost over probability distribution of malicious and benign domain names.

\section{Evaluation}
\label{evaluation}
The results of our experiments are divided into two sections. Firstly, we evaluate the performance of the newly proposed LSTM model and compare it with a previous work known as DeepDGA in terms of model accuracy \cite{anderson2016deepdga}. In addition, we report on how the model accuracy changes against tempering of input samples in order to generate adversarial instances. Finally, we analyze the DGA classifier before and after adversarial augmentation of training data. %uld Finally, we evaluate the performance of a PROSPECD container that is used to store blacklist of malicious domain names and relevant information. % on how and when the blacklist was created. 

At first, a binary classification, which simply predicts between DGA-based malicious or benign (Alexa top 135K) samples are applied. Table \ref{comparison} demonstrates a comparison between the performance of the Deep-DGA model and our LSTM-based model. The detection rates of Cryptolocker and Dicrypt is higher with DeepDGA than our DGA classifier with the available samples. On the other hand, Locky V2, Pykspa, Ramdo and Simda are detected with better accuracy by our classifier, and the rest of the cases show the same detection rate for both. Even though the results demonstrated that the improvement was not substantial, the results could have turned out to be different, since we used a different dataset than DeepDGA.

\begin{table}[!h]
\centering
\caption{Deep-DGA and the Proposed Model Comparison}
\sisetup{per-mode=symbol}
%\renewcommand\arraystretch{0.95}
%\setlength{\doublerulesep}{5pt}
%\rowcolors{3}{red!30}{cyan!5}

\begin{tabular}{||c c c||} 
 \hline
     & Deep-DGA & The Proposed DGA Detector \\ [0.5ex] 
 \hline\hline
 Corebot & 1.0 & 1.0 \\ 
 \hline
 Cryptolocker & \textbf{1.0} & 0.98 \\
 \hline
 Dircrypt & \textbf{0.99} & 0.98 \\ 
 \hline
 Locky V2 & 0.97 & \textbf{0.99} \\
 \hline
 Pykspa & 0.85 & \textbf{0.90} \\
 \hline
 Qakbot & 0.99 & 0.99 \\
 \hline
 Ramdo & 0.99 & \textbf{1.0} \\
 \hline
 Ramnit & 0.98 & 0.98 \\
 \hline
 Simda & 0.96 & \textbf{0.98} \\
 \hline
 Average & 0.97 & \textbf{0.98} \\
 \hline
 \end{tabular}
\label{comparison}
\end{table}

\begin{comment}
\begin{table}[!t]
    \centering
    \includegraphics[width=8cm\linewidth]{comparison_1.pdf}
    \caption{ Comparison Between the Deep-DGA and the Proposed Model.}
    \label{comparison}
\end{table}
\end{comment}

\subsection{ LSTM model results}

\begin{table*}
\begin{center}
\scalebox{.9}{
  \begin{minipage}[b]{0.5\textwidth}
  \begin{center}
    \includegraphics[width=\textwidth]{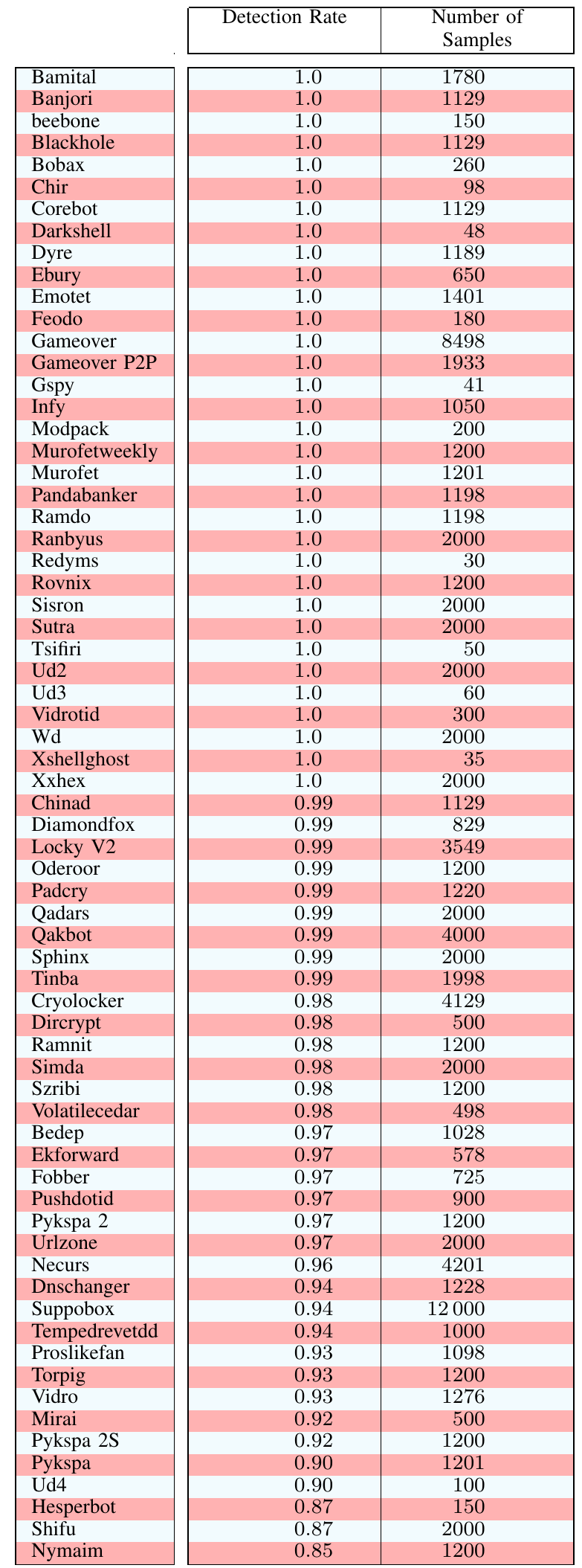}
    \caption{Detection Rate of DGA Malware Families Using the LSTM Model.}
    \label{fig:1}
    \end{center}
  \end{minipage}
  \begin{minipage}[b]{0.50\textwidth}
   \begin{center}
    \includegraphics[width=\textwidth]{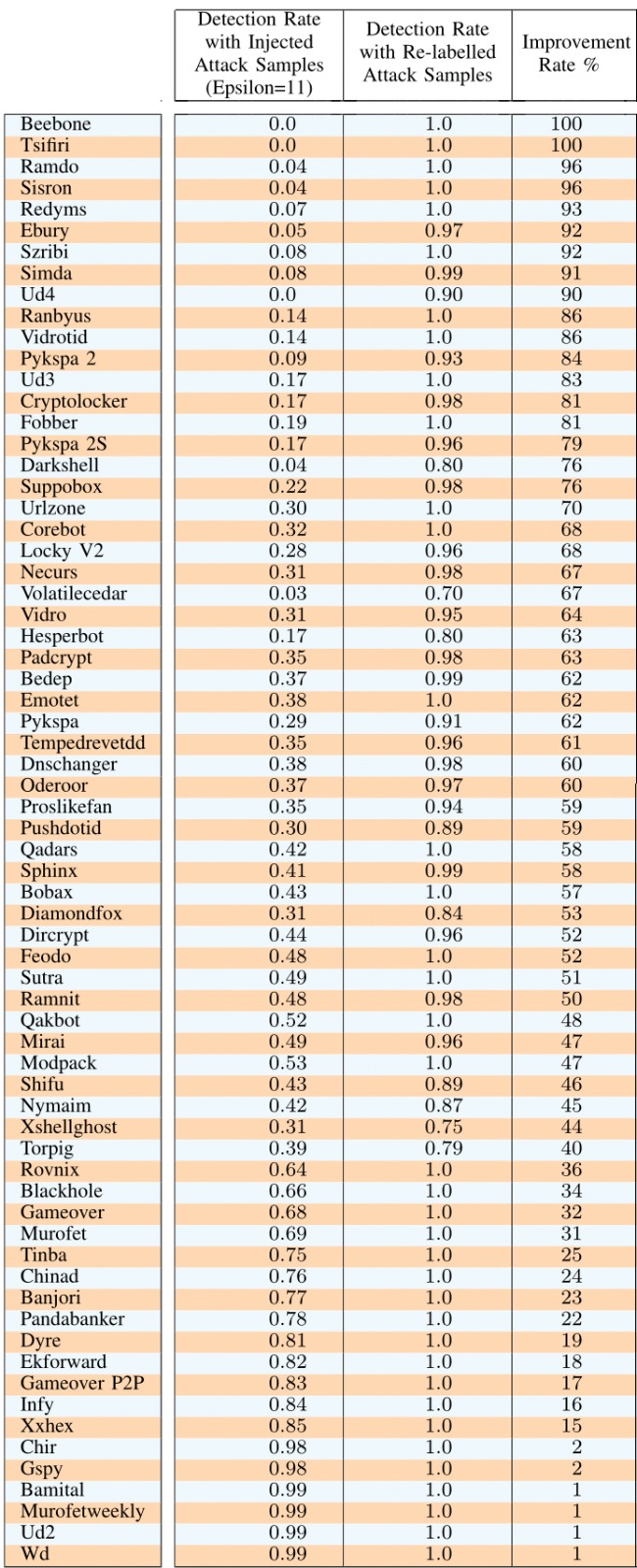}
    \caption{Detection Rate of DGA Malware Families Before and After Training Data Augmentation.}
    \label{fig:2}
     \end{center}
  \end{minipage}
  }
  \end{center}
\end{table*}

\begin{table*}[!ht]
\begin{center}
\includegraphics[width=12cm]{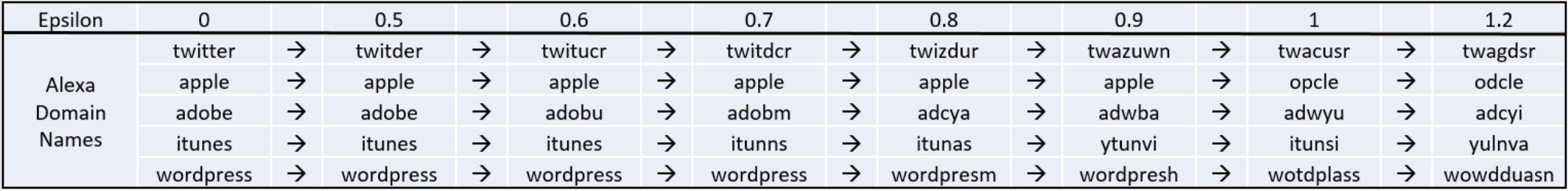}
\caption{Transformation of Alexa Domain Name Samples.}
\end{center}
\label{transformation}
\end{table*}

\begin{table*}[!ht]
\begin{center}
\includegraphics[width=11cm]{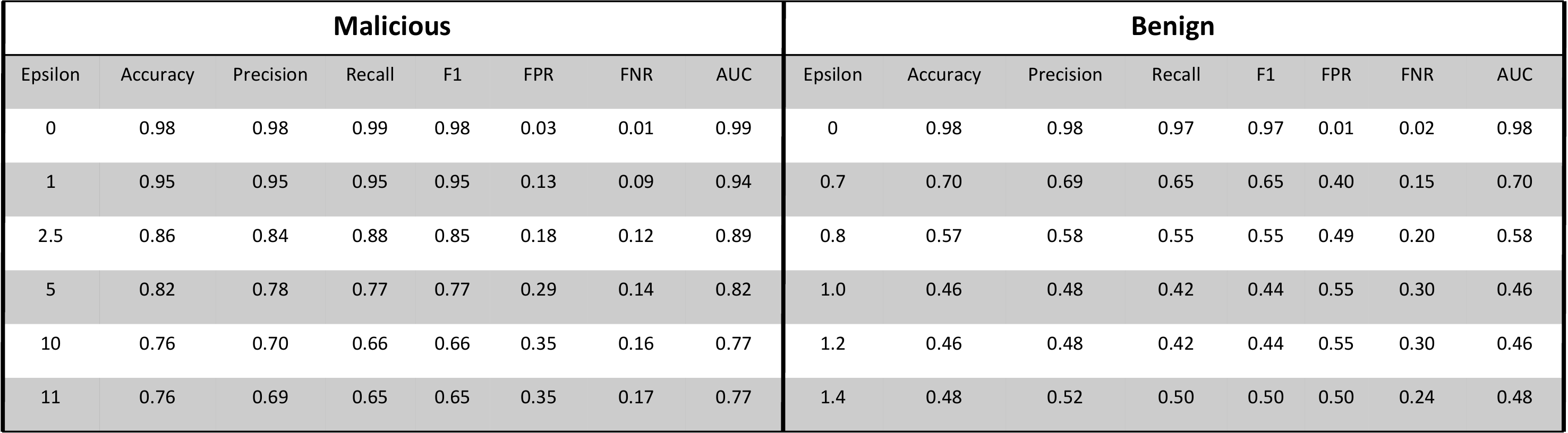}
\caption{Performance of the DGA Classifier vs. Penetration Coefficient for Both Benign and Malicious Domains.}
\end{center}
\label{accuracy}
\end{table*}

Table \ref{fig:1} shows the resulting detection rates for our model for 68 DGA families. Our findings show that our method performed with high accuracy (usually above the 0.97 accuracy margin) for most of the DGA-based malware families.

\begin{comment}
\begin{figure}[!ht]
\centering
\includegraphics[width=5.5cm]{figures/malicious2.eps}
\caption{Accuracy vs. Penetration Coefficient for Malicious Domain Names.}
\label{malicious}
\end{figure}

\begin{figure}[!ht]
\centering
\includegraphics[width=5.5cm]{figures/benign.eps}
\caption{Accuracy vs. Penetration Coefficient for Benign Domain Names.}
\label {benign}
\end{figure}
\end{comment}

We also evaluate the DGA classifier performance considering standard metrics such as precision, recall, F1-score, false positive rate (FPR), false negative rate (FNR) and area under the receiver operating characteristic curve (AUC). These evaluation metrics are widely used to measure the quality of the ML model. Using the proposed algorithm, we craft domain names from both benign (Alexa 135K) and DGA-based malicious samples. Based on different epsilon values, the changes in these evaluation metrics of the model can be viewed in Table VI. Initially, we set the epsilon value to zero to observe the actual performance of the model. Our findings show that the model performs well in terms of the aforementioned metrics. In case of the adversarial samples that were generated from malicious domain names, the accuracy rate of the LSTM model degrades with increase in epsilon value until it stabilizes at an equilibrium, because, at that point, the model has been trained well enough to recognize malicious domains. In addition, the dissimilarities between the benign class and the malicious class drastically increase. This indicates the limit of misclassification, even with increasing epsilon values.

\begin{comment}

Figure \ref{malicious} demonstrates how the addition of noise at different levels to the malicious samples helps to create the adversarial instances with the proposed algorithm. 
\end{comment}

We also consider benign instances as an input to corrupt the benign samples for creating the adversarial domain names. Subtle perturbations do not decrease an accuracy much, since the injected epsilon values do not manipulate the original data sufficiently to cause misclassification. When we continue to increase the penetration coefficient that causes slight differences to the original benign data, the model fails to recognize these changes. Therefore, the model performance is dramatically impaired. As we further scale up the noise, the model starts to predict these drastic modifications more accurately, due to severe degradation of the input. Table V shows how the domain name samples are transformed by the epsilon values. %The examples of these transformations due to the epsilon values are shown in Table V.

It is noteworthy to observe the decreasing accuracy of our DGA classifier as we add perturbations because adversarial examples mislead the model into making incorrect decisions that increase the number of false positives and false negatives. The various adversarial samples that are created  by injecting noise has the potential to deceive the LSTM model even more than the adversarial samples generated by GAN. In the study of Anderson et al. \cite{anderson2016deepdga}, it was found that the detection rate of the model is 48.0\%, which means that they achieved an attack success rate of about 53\%. Table VI shows instances of LSTM model's accuracy to be around 45\% with different perturbation coefficients. We achieved the highest attack success rate of 56\%, which is higher than the GAN approach by 3\%, indicating that our model generated DGA families are able to deceive the ML model more effectively.

\begin{comment}
\begin{table}[!ht]
\includegraphics[width
\caption{Accuracy vs. Penetration Coefficient Results for Both Benign and Malicious Domains.}
\label {accuracy}
\end{table}
\end{comment}

\subsection{Improving the LSTM Model with Augmented Training Data}
As discussed above, we are able to successfully produce adversarial domain names that can bypass detection by the LSTM model. We show that successful augmentation of training data samples can be done with our proposed method. Changes in penetration coefficients can impact the DGA classifier to different extents in terms of the model accuracy.

We later modify the dataset by injecting correctly labelled adversarial domains. We replace every malicious training samples with its adversarial counterpart including the top Alexa 135K in the training set, and re-train the model. Table \ref{fig:2} illustrates the differences before and after training with adversarial samples when the epsilon value is 11. Our reason for selecting the value 11 for the epsilon is to illustrate the maximum damage to the well-trained LSTM model and how training the model with augmented data performs much better.
 
When the model is trained with adversarial samples, the model is able to detect unseen malicious samples in the training set to a larger extent. The hardened classifier increases the model's detection ability for each DGA family, as can be seen from Table \ref{fig:2}. As noted for some family groups, such as Bamital, Gspy, and Ud2, the adversarial manipulation did not have any significant impact on model accuracy (within 1\%). However, for most others, the training with augmented data boosted accuracy immensely, on some occasions reaching up to 100\%. As a result, the model trained with adversarial samples has shown to perform much more accurately, close to the performance of the model before adversarial manipulation.

\section{Conclusions}
\label{conclusion}
In this paper, we presented a novel detection system based on an LSTM model for the detection of both traditional and dictionary-based DGA generated domains using the character-by-character approach. Our experimental findings show that the LSTM-based model can detect malicious domains with a trivial margin of error.

However, machine learning models are unable to learn the characteristic behaviors of DGA-based malicious domains if there are new or never-seen-before data in the testing dataset. In this study, we highlight this issue with an adversarial manipulation using different data perturbation cases. According to our findings, newly generated domains using the proposed perturbation approach could not be detected by the DGA classifier. After we trained the model with the augmented training dataset, including adversarial samples, the experimental results show that the LSTM model was able to detect previously unobserved DGA families.

We store malicious domain names, detected by our model, in a Protected Spreadsheet Container with Data (PROSPECD). It provides data confidentiality and integrity, %uld of detected malicious domain names
as well as origin integrity, role-based and attribute-based access control. % PROSPECD guarantees that an authenticated entity can access only those datasets for which the entity is authorized. 
% PROSPECD protects stored encrypted data in transit and at rest and, thus, protects the data against adversarial attacks.
PROSPECD protects the domain names in transit and at rest against adversarial access and modifications. %unauthorized modifications. 

\bibliographystyle{IEEEtran}
\bibliography{ibrahim
}

\end{document}